\documentclass[preprint,3p,twocolumn]{elsarticle}

\usepackage{natbib}
\usepackage{paralist}
\usepackage{rotating}
\usepackage{graphicx}
\usepackage[pdftex, colorlinks,
            linkcolor=black,
            citecolor=black,
            filecolor=black,
            urlcolor=black]{hyperref}
            
\usepackage{amssymb}
\usepackage{graphicx}

\usepackage{graphicx}
\usepackage[table]{xcolor} 
\usepackage{hyperref}
\usepackage[numbers]{natbib}
\usepackage{longtable}
\usepackage{paralist}
\usepackage{array}
\usepackage{booktabs}
\usepackage{multirow}
\usepackage{adjustbox}

\usepackage{framed}
\usepackage{pdflscape}
\usepackage[autostyle, english=british]{csquotes}
\usepackage{framed}

\journal{Journal of Systems and Software}

\begin{document}

\begin{frontmatter}



\title{Empirical Software Engineering: From Discipline to Interdiscipline}

\author[label1]{Daniel M\'{e}ndez Fern\'{a}ndez\corref{cor1}}
\ead{daniel.mendez@tum.de}
\author[label2]{Jan-Hendrik Passoth}

\address[label1]{Software and Systems Engineering, Technical University of Munich, Germany}

\address[label2]{Munich Center for Technology in Society, Technical University of Munich, Germany}

\cortext[cor1]{Corresponding author}

\begin{abstract}

Empirical software engineering has received much attention in recent years and coined the shift from a more design-science-driven engineering discipline to an insight-oriented, and theory-centric one. Yet, we still face many challenges, among which some increase the need for interdisciplinary research. This is especially true for the investigation of social, cultural and human-centric aspects of software engineering. Although we can already observe an increased recognition of the need for more interdisciplinary research in (empirical) software engineering, such research configurations come with challenges barely discussed from a scientific point of view. In this position paper, we critically reflect upon the epistemological setting of empirical software engineering and elaborate its configuration as an \emph{Interdiscipline}. In particular, we (1) elaborate a pragmatic view on empirical research for software engineering reflecting a cyclic process for knowledge creation, (2) motivate a path towards symmetrical interdisciplinary research, and (3) adopt five rules of thumb from other interdisciplinary collaborations in our field before concluding with new emerging challenges. This supports to elevate empirical software engineering from a developing discipline moving towards a paradigmatic stage of normal science to one that configures interdisciplinary teams and research methods symmetrically. 
\end{abstract}


\end{frontmatter}


\section{Introduction}
\label{sec:Introduction}

Starting as a byproduct in a hardware-dominated world, software has become the main driver for entire industries and a transformative power in many fields of contemporary society. Software engineering practice and research are likewise continuously evolving to cope with the emerging challenges imposed by its ubiquitous nature: practical, institutional, and cultural contexts of software are dynamic and in constant change, and as they influence the shape and direction of software development, boundaries between systems and application domains become fuzzy. Software engineering today typically takes place in settings where we need to address, inter alia, application domain-specific questions (e.g. on domain-specific terminologies, concepts, and procedures), ethical questions (e.g. moral assessments in context of safety-critical situations), juridical questions (e.g. on data privacy or regulations of algorithms and their environment respectively), psychological questions (e.g. on improvements of team communications or working environments), or social and political questions (e.g. on societal impacts of software-driven technologies, the concerns of heterogenous actors, or accountability issues). Human actors -- whether customers, end users, or developers -- and their interests, needs, and values, but also their cognitive capabilities, fears, experiences, and expertise render software development endeavours as something individual and unique rather than something standardised and strictly formalised. In the end, software is developed by human beings for human beings and what works in one organisational context might be completely alien to the culture and needs of the next.

This poses new challenges for configurations of actors, skills, and methods in research and practice, as well as on the education of future software engineers (and end users). Although we can already see more and more calls for more interdisciplinary research and the integration of non-technical skills in higher education (see, e.g.,~\cite{Mitchell05, STEMBER19911}), the calls and proposals -- especially those in Software Engineering -- concentrate largely on educational aspects, e.g. on how to reduce the gap between isolated disciplinary conditions in academia and multidisciplinary real-life conditions in practice (see, e.g.,~\cite{Jazayeri04, Kuhrmann2013TeachingSP}). \emph{Interdisciplinary research}, however, comes with challenges barely discussed from a scientific, epistemological point of view in the software engineering community. Fields like health care and medicine, biology and neuroscience, or education have explicitly tackled such issues in the last two decades, in parts driven by the need to reflect on conditions of success of NSF and EU funding initiatives to integrate \enquote{ethical, legal and social issues / aspects} (ELSI/ELSA). Here, scholars discussed and defined \enquote{multidisciplinary}, \enquote{interdisciplinary} and \enquote{transdisciplinary} research to classify the different challenges and opportunities that arise from such configurations ~\cite{ROSENFIELD19921343, bogner_inter_2010, maasen_transdisciplinarity_2006}:

\begin{compactitem}
\item \emph{Multidisciplinary} projects involve researchers from various disciplines addressing a common problem in parallel (or sequentially) from their disciplinary-specific bases integrated by assembling results in a common book or workshop.
\item \emph{Interdisciplinary} projects build on a collaboration of researchers working jointly, but each still from their disciplinary-specific basis, to address a common problem
\item \emph{Transdisciplinary} projects involve researchers, practitioners and actors from governance agencies, NGOs or companies trying to work out a shared conceptual framework instrumentalising concepts, approaches, and theories from their parent disciplines.
\end{compactitem}

Being involved in inter- and transdisciplinary projects in stem cell research, neuroscience or urban planning, scholars have argued that even interdisciplinary collaborations involving only researchers from two or three disciplines create challenges very close to those formerly discussed only in the case of transdisciplinary projects: without investing time and effort (and money) into the search for common problems, a provisional but common language, and institutional backup, interdisciplinary projects tend to turn into multidisciplinary ones. This holds true already for \enquote{close} interdisciplinary collaborations between fields like industrial automation and software engineering. More pressing, but also more rewarding research challenges, as we will argue, emerge when trying to integrate research on the social, cultural and human-centric practices and contexts of software engineering. Dealing with issues that arose from the proliferation of tools and frameworks and the complexity growth in software engineering projects, software engineering has already turned itself into an evidence-driven empirical discipline, commonly known as empirical software engineering. However, the emerging major challenges need far more symmetrical forms of inquiry and design going beyond not only the preference for rationalism and formal reasoning still dominant in large parts of software engineering research, but also beyond the forms of empiricism already in place. They give rise to the need of symmetrical interdisciplinary research configurations at eye level to come to valid, but still manageable solutions to the problem of balancing different epistemic and practical requirements and standards. 

Placing symmetrical interdisciplinary configurations at the heart of empirical software engineering research -- especially when involving social, cultural, and human-centric facets -- has effects on the way that software engineering as a field can address questions central to defining it as a scientific discipline, such as \emph{What qualifies as (good) scientific practice?}, \emph{What counts as theory, as methodology, and as evidence?} or \emph{How are scientific controversies opened up, embraced, and closed?} Like in other inter- and transdisciplinary configuration, many conceptual problems and methodological issues in our field cannot be organised in terms of one single and dominant epistemological framework if we do not want to close down necessary and fruitful exchanges. Because of the high diversity of socio-economic and technical factors that pervade software engineering environments, it is -- not only from a practical and pragmatic perspective, but also from an epistemological point of view -- not sufficient to just \emph{use methods and concepts} from various disciplines in our research, but we need to effectively \emph{integrate methods and research competences} from the various disciplines.

\textbf{Contribution.} In this position paper, we critically reflect upon the broader epistemological setting of empirical software engineering and elaborate its configuration as an \enquote{Interdiscipline}. We will argue for stopping to treat empirical software engineering as a developing discipline moving towards a paradigmatic stage of normal science, but as a configuration of interdisciplinary teams and research methods - an \emph{interdiscipline}. To this end, we will make the following contributions: We (1) elaborate a pragmatic view on empirical research for software engineering that reflects a cyclic process for knowledge creation, (2) motivate a path towards symmetrical interdisciplinary research, and (3) adopt five rules of thumb from other interdisciplinary collaborations. 

The key addressees of this manuscript are twofold: (1) Scholars who are new to empirical software engineering (or parts of it) in general and interdisciplinary research in particular, and (2) scholars already aware of the importance of empirical research methods and interdisciplinary research, and interested in a broader epistemological view. 

\textbf{Outline.} We will first briefly elaborate on the epistemological, methodological, and pragmatic background of research methods already used in empirical software engineering and highlight consequences for theory building, methods development, and application as well as for interdisciplinary collaborations (Sect.~\ref{sec:ScienceNutshell}). We use those insights to elaborate a pragmatic view on empirical research for software engineering that reflects a cyclic process for knowledge creation in Sect.~\ref{sec:PragmaticCycle}. In Sect.~\ref{sec:Challenges}, we will discuss currently pressing challenges in empirical software engineering, show how they are grounded in basic controversies, and conclude by outlining a roadmap to turn empirical software engineering into an interdiscipline in Sect.~\ref{sec:TurningESEIntoInterdiscipline}.

\section{Scientific Movements and Practices \\-- in a Nutshell}
\label{sec:ScienceNutshell}

In response to the emerging challenges in software engineering, we have already seen a significant turn towards empirical approaches. Various contributions have baptised the shift from a more design science-oriented discipline driven by the application of scientific concepts and methods to practical ends to a more insight-oriented and theory-centric discipline as \emph{empirical software engineering}~\cite{KDJ04, Rombach2013, Basili:2013:Perspectives}. This turn towards empiricism in software engineering research and practice can be understood merely as a surplus to formal design orientation and as an avenue to turn the conceptual, theoretical, and mathematical work in computer science into more easily adoptable tools for practical applications in software engineering projects \enquote{out of the lab, and into the wild}. It can, however, also be understood as an epistemic intervention to turn software engineering into a field dealing with empirical evidence and methodologically gained insights. 

The consolidation of an empirical software engineering community is not happening in a straight-forward and united manner, but as the establishment of various, sometimes competing, sub-communities: while initial debates focussed mainly on issues of \enquote{rationalism versus empiricism}, we can today still observe some separate camps under the common banner of empiricism, focussing on issues like \enquote{qualitative studies vs. quantitative studies} or \enquote{students vs. professionals} (as subjects)~\cite{FJW+18}. In this manuscript, we will not even try to address the social mechanisms within the research communities to the extent they deserve, but briefly reflect and reason about the epistemological setting to motivate the challenges in our still emerging field. An insightful introduction into the evolution of empirical software engineering can be taken from the personal perspective provided by Victor Basili, one of the pioneers in the field~\cite{Basili:2013:Perspectives}.

In the following, we briefly highlight important concepts and major historically grown, and often competing, views on science as a form of knowledge and practice. We do not intend to discuss approaches to the philosophy of science in detail, but only to the extent necessary for shedding light onto the ongoing debates and struggles in empirical software engineering in its attempt(s) to understand and position itself as a scientific endeavour. An excellent introduction into the philosophy of science and its evolution can be taken from~\cite{Chalmers:1982aa}.

\subsection{The Ghosts of Rationalism and Empiricism}

Depending on the weapons of choice regarding epistemology -- the theory of knowledge -- and ontology -- the study of what exists -- there are various attempts to define what the ultimate goal of science is and how it is best achieved. In the 17th century, when science as an institution was still young, members of the newly founded Royal Society were trying to define and defend a set of practices both different from seeking truth in religious texts and the alchemistic attempt to turn copper or iron into gold or quicksilver. Rationalists (or intellectualists) like Descartes or Hobbes argued that because there is no serious link between a \enquote{res extensa} -- a realm outside -- and the \enquote{res cogitans} of our minds ~\cite{descartes_discourse_1993}, the only way to valid knowledge is deducing universal truths from basic principles using reasoning by geometry and mathematics. Empiricists like Hume or Boyle argued on the contrary that certain, if not all scientific problems cannot be deduced from basic principles -- simply because we cannot prove the axioms and reasoning tends to favour logical speculation -- and that we need to look for empirical proofs~\cite{Hume:1739aa}. This turned the search for knowledge into a never ending process of induction, i.e. the attempt to infer general rules from particular cases. The struggle about the right way to institutionalise modern science was fierce: Hobbes and Boyle, for example, attacked each other not only intellectually, but also personally and politically ~\cite{shapin_leviathan_1985}. 

Rationalism and Empiricism both have very distinct consequences in respect to what counts as \enquote{theory} or \enquote{truth} and although no serious approach in the philosophy of science today would repeat the old arguments for or against either of the two, the controversy is revived again from time to time in the form of implicit assumptions of what counts as \enquote{good}, \enquote{sound}, or \enquote{rigorous} science. Some of the arguments for or against empirical evidence in many engineering fields today still follow these old lines of demarcation (see also Sect.~\ref{sec:Challenges}).

\subsection{The Rise and Fall of the Unity of Science}
\label{sec:RiseAndFallOfUnity}

At the beginning of the past century, an informal group in Vienna that started to discuss such diverse works as the writings of Duhem, Lenin, and Frege served as a breeding ground to some of the most pressing problems in the philosophy of science of that time. The so-called \emph{Vienna Circle} tried to rework some of the basic premises of empiricism to link them with modern logic and therefore, eventually, overcome the differences between inductive and deductive strategies. Some of its suggestions include the use of so-called \enquote{protocol sentences} to standardise the translation of sensory data into reasonable input for logical calculus~\cite{Carnap:1932aa}, i.e. the use of verification as a principle to distinguish the scientific use of language from its misleading natural language sibling or the ultimate search for a \enquote{unity of science}–\cite{Carnap:1934aa, Neurath_1938} where any concept, theory, and statement of any potential disciplinary background is both empirically grounded and formulated in a common framework. One might say those suggestions were all not successful, but they failed in fact so big that their ghosts still haunt us today. The idea of a unified science, for example a science with one language (mathematics, logic, and verified statements about observations in those rare cases where mathematics and logic are simply too complicated), still hinders fruitful exchange in interdisciplinary projects -- both in cases of close disciplines such as theoretical and experimental physics and in cases of projects that try to connect knowledge from domains further apart. 

Two of the biggest critical publications refuting some of the main arguments of Logical Positivism -- Karl Popper's \enquote{Logik der Forschung} (The Logic of Scientific Discovery)~\cite{Popper:1959aa} and Thomas Kuhn's \enquote{The Structure of Scientific Revolutions}~\cite{Kuhn:1962aa} -- were published in the Vienna Circle's monograph series. Popper famously argued against verification and for falsification of hypotheses\footnote{When building and evaluating theories, we cannot test the theories themselves, but their consequences via hypotheses which are statements proposing a suitable explanation of some empirical phenomena~\cite{Stanley14}.} as a hallmark of the validity of a truth claim: No matter how hard we try, we can never verify a statement by inductive reasoning, but we can falsify it when a single observation does not fit our general rule (cf. null hypothesis testing). Kuhn, on the other hand, argued that it is only true for some time periods that science works the way that Popper said: problem- and puzzle-solving, conservatively sticking to theories, methods and statements that have not yet been falsified. But even in the \enquote{hardest} of all sciences (physics, chemistry or astronomy), history is full of revolutionary overturns, where paradigms shift, believes radically change, and where whatever was understood as true had to be evaluated once more in the light of a new paradigm. According to both, any scientific statement is only true as long as it fits the current paradigm -- a socially accepted system of believes, preferred procedures, and unquestioned doctrines in a given field  -- and as long as it is not critically refuted by an empirical observation that, again, is only valid in the light of a current paradigm.

\subsection{Collaborations and Science Wars}

Progress in science can only move forward through empirical observations as there are no other options for falsification. However, both the validity of theory and observation are bound to the social and cultural mechanisms of approval and acceptance of a paradigm by respected peers that are in many cases more related to judgements of \enquote{novelty}, \enquote{relevance}, or even \enquote{aesthetics} (considering, e.g., mathematical work). This might not cause problems in coherent peer groups working in a \enquote{normal science} mode under a shared paradigm, but it can cause massive controversies in times of paradigm shifts in one field and even more chaos when paradigms from different fields collide. 

During the 1970's and 1980's, such collisions caused a massive outrage on both sides when anthropologists, historians, linguists, semioticians and sociologists started conceptually and empirically to treat the practices, organisational mechanisms, and cultural patterns in scientific laboratories \enquote{symmetrically}~\cite{bloor_knowledge_1976, callon_elements_1986}: they observed, analysed, and interpreted the day-to-day work of science (experimenting, testing, writing, arguing, creating careers) like they treated any other field of practice in modern societies (families, political parties, market-based trading, \ldots)~\cite{latour_laboratory_1979, knorr-cetina_epistemic_1999}. This caused what became known as the \enquote{Science Wars}, in which each side accused the other of ignorance, stubbornness, and even hostility~\cite{sokal_impostures_1997,latour_you_2003}.

It is reasonable to say that the struggles were unproductive and unnecessary, to say the least, and a closer look at their historical background shows that they were mainly fueled by material and status interests that instrumentalised epistemological, methodological, and ontological arguments. And although the last three decades have seen various productive and insightful collaborations between scholars from very different backgrounds -- anthropologists and biologists working together on life science issues, historians and geologists working together on understanding the effects of human intervention into ecosystems, sociologists and computer scientists working together on Human-Computer-Interaction problems or critical algorithm design -- some resemblances of the science wars can still be observed from time to time when issues of authority (or funding) are involved.
 
\subsection{Theory, Research, and Evaluation as Practice}
\label{sec:TheoryBuilding}

The most fruitful collaborations between scholars from various backgrounds build on a very simple, but particularly hard to accept insight: Science -- the construction, testing, and evaluation of theories, models, arguments, experiments, and evidence -- is, if taken seriously, not an abstract and universal search for truth. It is a set of practices, institutions, and processes structured by changing rules, norms, and paradigms. In fact, there is nothing absolute about truth and there is no such thing as a universal way of scientific practice~\cite{McComas98}. What counts as a valid claim, a rigid approach, or a justified belief is relative to what counts as the shared standards and interpretations of validity, rigidity, or justification in the respective fields and subfields of scientific practice. Concepts and empirical methods have very specific purposes -- exploratory or explanatory, oriented towards depth or overviews, towards theory-generating or theory-testing -- each relying on different forms of empirical data -- qualitative or quantitative, procedural or categorical, observational or transactional. Such an understanding of science as a practice (and culture)~\cite{pickering_science_1992} has been corroborated both in the field of \enquote{Science \& Technology Studies}, an interdisciplinary field connecting engineers and philosophers, mathematicians and historians, computer scientists and sociologists to empirically study how science and the development of technologies is actually done in labs, offices, or at conferences~\cite{felt_handbook_2017} and by arguments developed by pragmatist philosophers of science. 

Following the early works of John Dewey~\cite{Dewey1938-DEWLTT} and Charles S. Peirce~\cite{Peirce2009-PEITNO} from the early 20th century, pragmatist thinkers and empirical researchers have argued that even if it is not possible to develop and justify a gold standard for what counts as pure and well defined scientific practice, one can work with something like a least common denominator. Dewey famously argued that any search for knowledge is following a common and ever repeating process of taking something for granted, experiencing an unsettlement of that belief, search for different hypotheses, and experimenting in order to test hypotheses before re-settling a new belief. While classic rationalists argued for \emph{deduction} as a basic principle of science -- an application of a general rule to a particular case by inferring a specific result -- and classic empiricists argued for \emph{induction} -- the inference of a general rule from a collection of particular cases --, pragmatists argue for a constant back and forth. As long as there is no particular reason for doubt, the search for knowledge works by deductively applying explicitly or implicitly known rules. But if that application of taken for granted beliefs fails, the search for knowledge switches to an inductive mode until the situation is settled again and deduction is again possible. Peirce formalised that pragmatic insight in his works on logic and inference and proposed that there is a more hypothetical form of inference linking both -- an inference that he later called \emph{abduction} -- an educated, but hypothetical guess of the best explanation. 

Summarising the resulting three principles:
\begin{compactitem}
\item \emph{Induction} describes the inference of a general rule from a particular case: If one selects a sample of beans from a bag (case) and all selected beans are white (result), then all beans from this bag must be white (rule).
\item \emph{Deduction} describes the application of a general rule to a particular case: If all beans from a bag are white (rule) and a particular bean is from the bag (case) then this bean will be white (result).
\item \emph{Abduction} describes the (creative) synthesis of a case from a general rule and a particular result: if all beans from a bag are white (rule) and there is a handful of white beans available (result), then this handful is (probably) from the bag.
\end{compactitem}

If taken as a practice moving back and forth through the application of these three principles, we can connect them in the form of an \enquote{empirical life cycle} introduced next. 

\section{Pragmatic Cycle for Empirical Research}
\label{sec:PragmaticCycle}
Methodologies that are built upon pragmatist arguments have already shown that at least qualitative research has to be understood as an iterative process of generating empirically grounded theories from a repetition of induction, abduction, and deduction until a useful theoretical maturity has been reached. 

\begin{figure}[!ht]
\centering
\includegraphics[width=\columnwidth]{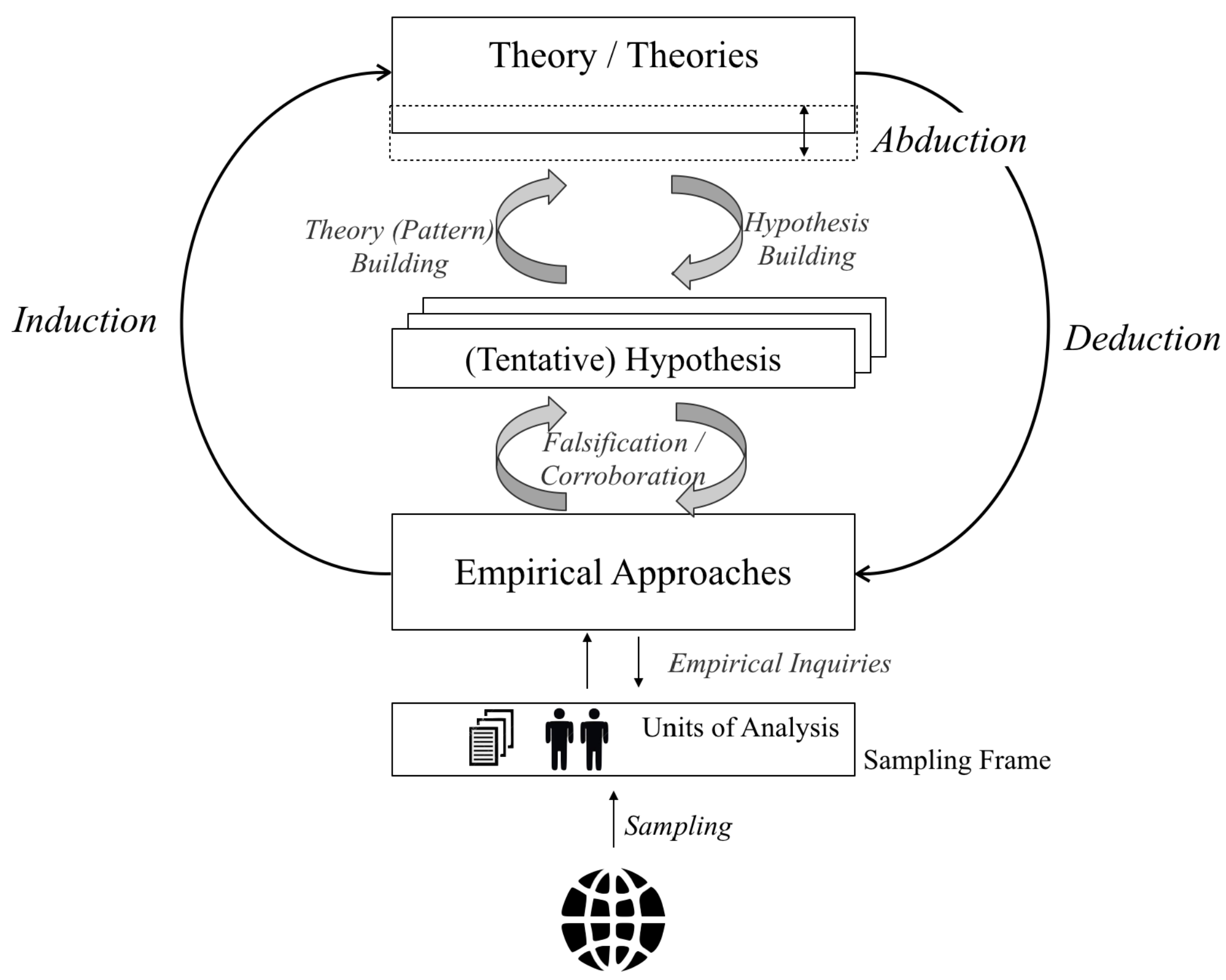}
\caption{Pragmatic cycle for empirical research.}
\label{fig:empiricism_nutshell}
\end{figure}

We now extend that idea in our attempt to capture an empirical life cycle that visualises knowledge creation in general as an iterative theory building and theory evaluation process as shown in Fig.~\ref{fig:empiricism_nutshell}. This process captures and combines the different concepts and principles introduced in the previous section yielding a pragmatic life cycle for empirical research such as in software engineering.
 
Our empirical life cycle recognises that there is nothing absolute about truth and that there no such thing as one universal way of scientific practice, but that there are many different and valid (and complementary) ways of undertaking research. There is no single, empirically inquired point of view that will ever provide us with an entire picture when interpreting relevant phenomena. 

Considering, for example, techniques to elicit software requirements, effectively testing the sensitivity of these techniques to practical contexts and building robust theories would require a combination of:

\begin{compactenum}
\item (Inductive) observational studies to explore practical real-life contexts and relevant parameters (such as constraints and objectives),
\item building concepts in relation to existing theories that consider both system-theoretical modelling aspects and socio-economic aspects alike (such as expectations on how specific techniques would yield improvements),
\item (deductive) quasi-controlled experiments to corroborate concepts by testing the expected improvements, and
\item scaling up to practice via in-vivo studies and replications to increase the robustness of the emerging theory.
\end{compactenum}

This example alone shows that even if considering theory building as the pragmatic iterative cycle of constant theory building and evaluation it is, it is still not a straight-forward one. In fact, there exist multiple paths through our lifecycle by applying different methods in different orders (e.g. from quantitative to qualitative investigations or vice-versa). Every empirical method, may it be a survey, a case study, or a controlled experiment, has its place in this lifecycle and the appropriateness of the method chosen and the underlying data type (qualitative or quantitative) depends on the research question asked. Theory building and evaluation is an iterative practice where theories are revealed, set into context to existing evidence and concepts, and continuously ``tested'' to refine them and increase their robustness and validity. 

\section{Challenges in Empirical Software Engineering}
\label{sec:Challenges}

Although the importance of empirical approaches in software engineering research and practice is already acknowledged, our still young field faces various challenges when it comes to provide robust scientific theories.  The interdisciplinary nature, the uncertainties of decisions on requirements or criteria for evaluating quality or usefulness, the dependencies of domains and the heterogenous stakeholders involved, and the growing complexity of software development processes open up very different avenues for empirical investigations ranging from mining software repositories to interviewing stakeholders or conducting in-depth case studies. The closeness of software engineering phenomena to the judgment by the individuals involved therein makes it inherently difficult to build general principles and theories; central questions are often heavily dependent on a socio-economic context where the boundaries are difficult to capture because of the interrelated human, economic, technological, and also cultural factors involved. 

As an analogy for highlighting the difficulties and limitations in theory building, a comparison with other evidence-based disciplines, such as medical research, can be drawn~\cite{KDJ04}. There, large sample sizes and constant replications are generally accepted for proper (statistical) conclusions while this is exactly what is so difficult in software engineering research. In software engineering, it is not only difficult to draw large sample sizes\footnote{At least when considering larger units of analysis such as ``Software Projects''.}, but also to reason about their representativeness. We have, in fact, very limited knowledge about true populations and their characteristics, especially when it comes to used methods, technologies, and development processes and their stakeholders: How many requirements engineers do generally exist? How many in Germany? What are their responsibilities and their levels of expertise? A further problem, finally, consists of that knowledge acquisition in software engineering is yet not always in tune with the pace of changes in our cases under investigation (e.g., industrial practices)~\cite{7203018}.

All this often feeds critical voices who reject the value of empirical research at all postulating, for example, that ``our inability to carry out truly scientific experiments [...] will yield anecdotes of limited value''~\cite{Parnas2003}. However, while we  agree that outside the flavour of pure mathematics -- which is often preferred by those opposing empiricism as the basis for theory building -- there is no certainty at all, it is empiricism that allows us to advance as a community to better reason about software engineering practices under realistic conditions. Software engineering always takes place under very specific and complex conditions that involve contexts, processes, and practices that, even if in some cases at least approximately expressible in mathematical terms, as in the case of economic indicators, cannot be integrated in a unified model. If the ultimately failed search for a unity of science (Sect.~\ref{sec:RiseAndFallOfUnity}) has taught us anything, then it is the epistemological and pragmatic insight that a standardised language -- of mathematics, of modern logic, of \enquote{truly} comparable protocols -- cannot be the gold standard for knowledge production. While we agree that the narrative and interpretative accounts created by empirical research have always only limited validity until a next iteration of our empirical life cycle comes up with refutations or specifications, they are the only way to bring software engineering research closer to the context and conditions for which software is developed and in which software is used. 

\subsection{The State of Empirical Evidence}
Although recent developments have improved our empirical understanding of software engineering practices and processes, the current state of evidence is still weak when compared to other more mature fields. A large extent of our everyday practice in software engineering is still governed more by conventional wisdom than it is governed by empirical evidence. This is especially true for the social, cultural, and political aspects of software engineering, such as early stages of development, rendering the inference of robust theories inherently problematic. 

Even though we can observe an increase of empirical studies in the various fields of software engineering research, many studies still do provide either circumstantial evidence by focusing on isolated contexts without taking into account the relation to existing evidence or -- worse -- they neglect the context completely. The effects are portrayed by Jacobson's observation in context of the SEMAT initiative~\cite{jacobson2009semat}: software engineering is gravely hampered by (1) the prevalence of fads more typical of fashion industry than of an engineering discipline; (2) the lack of a sound, widely accepted theoretical basis; (3) the huge number of methods and method variants, with differences little understood and artificially magnified; (4) the lack of credible experimental evaluation and validation; and finally (5) the split between industry practice and academic research. The consequence of the current situation are best described by Wohlin et al. saying that ``there exists no generally accepted theory in software engineering [...]. Some laws\footnote{We consider a law to be a purely descriptive, analytical theory about phenomena without explanations for the phenomena.}, hypotheses and conjectures exist, but yet no generally accepted theory''~\cite{WOHLIN2015229}. As a matter of fact, a large extent of the theories (or theory patterns) we have for software engineering are still transferred from theories in other disciplines (e.g. organisational psychology), sometimes by adopting them, but mostly by transferring them verbatim~\cite{Sjoberg:2007:FEM:1253532.1254730}.

Software engineering itself however is often still governed by folklore turned into facts~\cite{Bossavit2015}. Similarly as in other fields before, many theories specific to software engineering emerged from the early times of the discipline where empiricism had no significance at all and where claims by authorities where often treated as facts. One prominent example for such a ``fact'' is grounded in the well-known essay by Edsger Dijkstra \emph{Go To Statement Considered Harmful}~\cite{Dijkstra68} from 1968, largely based on reasoning by argument and triggering a public exchange between different scholars via published notes (all considering the previous note as ``harmful'' itself). Although this exchange fostered an important and fruitful debate in the community at that time, it still remained largely a public exchange between scholars based on reasoning by argument. This did not change until 2015, nearly 50 years lager, when Nagappan et al.~\cite{NRY+15} published the results of their large-scale study analysing C code from GitHub repositories and suggesting that the use of goto statements in practice does \emph{not} appear to be harmful.

Despite the positive developments towards empirical research, software engineering claims are still often judged based on the number, faith, and vocal energy of their protagonists, and ``facts'' are often taken for granted based on dogmatic statements by authorities rather than based on empirical evidence. Exemplary symptoms of the resulting folklore are illustrated in Tab.~\ref{tab:Folklore}. 

Folklore and its grounding in common sense and conventional wisdom is, if understood from a pragmatist point of view, a form of taken-for-granted knowledge that is routinely used to frame known and unknown situations alike. But unlike the provisional beliefs that govern the empirical life cycle of scientific practice, folklore is backed up by so many different social and political mechanisms (dogma, influence of its protagonists, or the search for fame, to name just a few), that it is not doubted when an empirical observation does not fit. Even worse: it is this dismissal of unfitting empirical evidence in accounts that use folklore that as an effect fuels the troubling belief that empirical evidence in general is not able to cast doubt on what is already \enquote{commonly known}.

\begin{table*}[htbp]
\caption{Exemplary Folklore in Software Engineering}
\label{tab:Folklore}
\begin{center}
\begin{tabular}{|p{16cm}|}\hline
\footnotesize
\textbf{Example: Goal-oriented RE.} One example for the low state of empirical evidence can be seen in the field of goal-oriented requirements engineering (GORE) with a dominance of GORE solution proposals not reflected in industrial everyday practice. Horkoff et al.~\cite{HAC+16} analysed in a systematic literature review, including 966 papers published by 2016, 246 papers in detail. Out of these, 131 indicate to include a case study. While at first sight maybe positively surprising, a subsequent study by Mavin et al.~\cite{Mavin+17} revealed that only 20 of these 131 case studies were actually conducted in practical in-vivo settings. In contrast, a large-scale investigation on the status quo and problems in RE indicates to that roughly 5\% of industrial environments rely on this RE technique at all (see also \url{www.napire.org}). While this does certainly not imply that the practical relevance of the contributions is weak, it implies that we have little knowledge about their relevance and impact, let alone their practical conditions and effects. We could argue that many of these contributions provide solutions to problems not yet well understood. \\

\footnotesize
\textbf{Example: Chaos Report.} It is difficult to mayor in software engineering without learning about the Chaos report figures by the Standish group providing figures on reasons for software failures and, in particular, cost overruns. While those figures are still frequently cited, there exist many reports already (e.g.~\cite{EV10, JORGENSEN2006297}) providing sufficient evidence-based arguments to doubt the validity of the Chaos Report and raise questions on major methodological flaws in how the figures were revealed. As the Standish Group never disclosed the data, a more accurate analysis (or replication) is, however not possible; yet it remains frequently used by researchers and (e.g. policy) advisors to support their own claims and hopes.\\

\footnotesize
\textbf{Causes and Effects.} Both examples could not be more different and yet they share similar causes and effects. As an empirical (cross-cutting) community, we are still far from a common ground with standards that hold among the various software engineering (sub-)communities and cultures. This fuels research projects isolated from practical contexts, thus, yielding conventional wisdom rather than evidence and creating, in turn, more folklore which, when being cited just often enough, at some point becomes accepted as universally true in the communities (remaining then difficult to eradicate).\\

\hline
\end{tabular}
\end{center}
\label{tab}
\end{table*}

\subsection{From Conventional Wisdom to Evidence}

If software engineering research wants to overcome this grounding in conventional wisdom and universal theories, then a change towards evidence-based, theory-centric and context-sensitive software engineering research is necessary~\cite{8048656}. This has been more and more understood and accepted over the last years and the growing empirical software engineering research community is fostering, in fact, great progress in methodological and scientific rigour~\cite{RALPH201868}, also by establishing standardised approaches and method guidelines to empirical software engineering (ranging from systematic mapping studies~\cite{PFMM08} to case study research~\cite{runeson09}). Yet, the path towards empiricism and robust theories in software engineering remains stoney as there is still a lot of uncertainty and confusion about the choice of appropriate configurations of research methods and about how to align new results with existing evidence -- a hallmark of any insight-oriented scientific practice. 

The trouble with building a reliable body of knowledge in any form of scientific practice is that there is no simple solution to it. There exists a variety of causes rooted in the lack of...
\begin{compactitem}
\item ...(methodological) awareness in less empirically oriented software engineering sub-fields
\item ...appreciation of qualitative studies even in empirically oriented software engineering sub-fields
\item ...empirical data disclosure (from study protocols to the actually analysed data itself)
\item ...transparency and replicability often emerging from the lack of (open) data 
\item ...appreciating replication studies and the reporting of null results
\item ...recognising context-specific conditions and the relation to existing evidence 
\item ...adoption of empirical research methods from other disciplines to software engineering contexts
\end{compactitem}

Although there is an increased awareness among scholars from various software engineering research (sub-)communities, only addressing as many of these issues as possible can yield a fruitful path towards robust software engineering theories and concepts. Otherwise, the theories we reveal remain context-independent and universal and, thus, not applicable to specific context situations as they would require folklore, conventional wisdom or educated guessing to make them work in specific cases. In those cases they are useful, they still study too often observable, quantifiable effects only (``what is happening?'') rather than providing explanations for core mechanisms in the phenomena (``why is it happening?''), as pointed out by J{\o}rgensen in his keynote to the 12th International Symposium on Empirical Software Engineering and Measurement~\cite{Jorgensen18}. 

As already shown in Sect.~\ref{sec:PragmaticCycle}, it is simply impossible to build robust theories with single-method studies alone or by relying on a specific data type only. Theory building and evaluation is a continuous iterative process relying on a variety of different empirical methods and each individual inquiry needs careful consideration of context factors and the relation to existing evidence. Otherwise, all we reveal are universal theories that are built upon apodictic arguments when what we need first is a solid basis with context-specific theories and explanations having survived multiple attempts for refutation. 

The path towards interdisciplinary research is surely not the sole measure of success to solve all the issues discussed above. The research community needs to further work on shared and accepted practices for empirical studies, for replication studies, and for data sharing, and it needs to further disseminate those practices into other software engineering communities. As mentioned earlier, we are already making great progress in this direction reflected, for example, in the increase of methodological guidelines to empirical research (see, e.g., ~\cite{runeson09,PFMM08,petersen2015guidelines,kuhrmann2017pragmatic,Stol+16,jedlitschka2008reporting}, just to name a few selected ones) or in open science policies and initiatives to recognise shared data sets becoming more and more prominent in software engineering venues (see, e.g., the open science policies established in recent empirical software engineering conferences and journals). 

It is, however, still far from reasonable to believe that we can yield robust software engineering theories by relying on one paradigmatic disciplinary framework only and by forcing every other approach into its rigid form. 

\subsection{From Discipline to Interdiscipline}

When it comes to integrating different forms of knowledge and evidence from other disciplines and subfields, a strategy under the umbrella of one sole epistemological framework has shown to be not very promising. This is already true in the case of close interdisciplinary cooperations (integration of, for example, different fields of engineering), but is even less viable when it comes to making use of evidence, concepts, and methods used in fields and disciplines that focus on more human-centric issues. Such an integration of multiple disciplines is necessary and has already been recognised as the only effective way to reach a more realistic understanding of human involvement in software engineering endeavours~\cite{Lenberg2015behavioral}. This human involvement should not be treated as a simple facet that plays a minor role, because it eventually characterises largely how whole areas of our discipline (and with them the use of approaches, methods, and tools) manifest themselves in practice; for instance:
\begin{compactitem}
\item early stages of development being sensitive to the particularities of domains, organisational contexts, and the various stakeholders involved (each with own hopes, desires, and beliefs),
\item software process models that specifically aim at team cultures and values (e.g., agile development),
\item software project organisation and management topics including planning and effort estimation, 
\item Security Engineering, in particular those aspects having a human-centric dimension (e.g. social engineering), or
\item more generally, human-centred engineering (as inherent to, e.g., HCI).
\end{compactitem}

Building concepts and theories that effectively include the social, cultural or, in general, human aspects (cognitive capabilities, belief systems, values) cannot be addressed by pure rational reasoning -- human practices, institutions, and values are purely rational in very rare cases only, if at all~\cite{Lambert06}. However, it can also not be effectively addressed within one sole epistemological framework, but only via a combination of methods and approaches (and skills). The use of single empirical methods is not sufficient anymore, but we need to triangulate by means of using various methods in combination. 

\section{Turning Empirical Software Engineering into an Interdiscipline}
\label{sec:TurningESEIntoInterdiscipline}

Addressing the emerging, especially human-centric challenges in software engineering research needs both recognising and addressing all challenges described above as they affect the effectivity of all forms of scientific practice, as well as establishing new forms of empirical inquiry that go beyond the capacities of a sole epistemological framework. The latter can only be reached with more \emph{symmetrical interdisciplinary research configurations} at eye level to come to valid, but still manageable solutions to the problem of balancing different epistemic standards. In the following, we elaborate on these forms of collaborations before concluding with some new challenges they bring.

\subsection{Establishment of Symmetrical Collaborations}

When transferring approaches, concepts, and methods from other disciplines, we not only adopt their application, but also the underlying theories. For instance, when employing qualitative methods like interviews or observations to explore practical contexts and social and cultural mechanisms involved therein, we do not only rely on a specific technique, but also on existing theories from social science, organisational studies, or psychology. For software engineering practice and research alike there barely exists such thing as greenfield engineering anymore. Existing theories from related fields and disciplines need to be considered and carefully set in relation to each other. In our requirements engineering example above, if taking it to an extreme, even system-theoretical modeling concepts and social theories need to be carefully set in relation to each other as both include structural mechanisms each with similar or even overlapping concepts (e.g. regarding the notion of ``interaction'' or ``trust''). It is naive at best and ignorant at worst to believe that a balanced combination of expertise can be achieved by research team configurations stemming from one discipline alone. However, the history of science, as discussed in Sect.~\ref{sec:RiseAndFallOfUnity}, has also shown that aiming high and for a unification of concepts, methods, and approaches for inter- and transdisciplinary research is also not very effective. 

The most pragmatic way towards a balanced combination of methods and domain expertise, we argue, is to set up flexible interdisciplinary collaborations and to shift from treating empirical software engineering as one singular discipline but as an \enquote{interdiscipline} where social, cultural, and human-centric issues shape the configurations of questions, research methods, and teams in the same way as mathematical models and procedural guidelines do already. We already argued that scholars from \emph{Science \& Technology Studies} -- an interdiscipline integrating actors, concepts, and methods from social sciences, humanities, natural sciences, and engineering  to study the various relations between science, technology, and society -- have developed forms of in-depth collaborations mainly in the life sciences, health, or biology that go beyond just adding \enquote{ethical, legal and social aspects} to projects and consortiums in these fields. 

We believe that these forms of collaboration can also be very helpful for empirical software engineering research and practice especially when tackling so-called \enquote{wicked problems}~\cite{Rittel_1973,Churchman_1967,CW97}. As Liegl et al. argue in the case of integrating ethics -- one field of wicked problems -- into the design of emergency response systems, such an approach is also in tune with the outlines of the current E.U. \enquote{Horizon 2020} funding framework, but needs experimental designs for collaborations, stakeholder integration as well as time and effort to treat the ethical and social implications as an "object of collective and ongoing negotiation, which needs to be done in situ and hand in hand with end-users and other stakeholders"~\cite{liegl_ethically_2015}. Balmer et al. proposed five general rules of thumb for such \enquote{post-elsi} interdisciplinary collaborations in general (and not only focussing on ethics, but on all kind of human-centric issues)~\cite{balmer_five_2016}, and which we can thus adopt to interdisciplinary empirical software engineering:

\begin{compactenum}
\item \emph{Collaborative experimentation}: The configuration of methods and approaches in specific projects need to be worked out experimentally and collaboratively. Instead of just plugging in some methods and concepts from one discipline last minute in an already designed project of the other discipline, the setup of research questions and respective methods should be worked out together and it is advisable to carefully plan in advance for finding a common, provisional language.
\item \emph{Taking risks}: Interdisciplinary collaborations involve risks, both professionally and personally. Major publication outlets and funding programs are and will be in the near future still organised on disciplinary terms leaving interdisciplinary journals, conferences, and funding schemes still the exception. Getting collaborative works published is difficult, to say the least, given the differences in paradigmatic standards or even in writing styles. This is true also personally in terms of career development. But we have ourselves experienced it to be worth the risks: engaging in collaborations early takes time, but it makes engaging in innovative and interesting projects later easier.
\item \emph{Collaborative reflexivity}: Reflexivity -- the application of a scientific standard and the empirical life cycle repeatedly to ones own practice, concepts, and methods -- is the only way to move away from intellectual stubbornness and sticking with folklore and conventional wisdom. But it can turn into a dangerous ally, leading to infinite navel-gazing~\cite{passoth_beware_2013}. Collaborative reflexivity -- confronting each other both with insights and concepts as well as prejudices and biases -- is challenging, but a very effective way to move forward.
\item \emph{Opening-up discussions of unshared goals}: In interdisciplinary collaborations, there are various individual and disciplinary expectations about the goals of the collaboration. Such collaborations will lead to more responsible innovation or better economic and organisational fit. Those expectations, however, are quite often not completely shared personally and judged against the standards of the respective home disciplines. Integrating time slots and spaces for open debate about unshared goals helps in this regard.
\item \emph{Neighbourliness}: 
There is no need for all participants in interdisciplinary collaborations to share all standards and practices and to strive for becoming alike. Treating the other disciplines as neighbours and as equals means engaging in boundary work~\cite{star_institutional_1989,gieryn_boundary-work_1983} and in searching for common objects, artefacts, and concepts that can also be used in individual disciplinary ways.
\end{compactenum}

\subsection{The Rise of New Challenges}
To turn empirical software engineering into an interdiscipline, we argued so far that we need to engage in coalitions with experts from other disciplines, such as social science or psychology, instead of treating each other as mere byproducts. This affects the design of multi-method programmes and the explicit consideration of theories and concepts from other disciplines instead of building various isolated new ones. More precisely, given our current state of evidence, we believe that it is important to concentrate on context-specific concepts and explanations as a basis for more useful and viable theories in the future that embed and link existing theories from various disciplines instead of just transferring them verbatim from the other disciplines.

Establishing this new perspective on empirical software engineering as an interdiscipline poses, however, new challenges such as:
\begin{compactenum}
\item When can and should existing theories from other disciplines be transferred (verbatim) to the software engineering body of knowledge, when should they be effectively integrated in parts only, and when do they need to be adapted?
\item Where and how do established method standards need to be modified to enable the effective design of manageable-sized short-term studies as well as long-term interdisciplinary research collaborations?
\item How can we effectively disseminate already exiting method standards across the existing software engineering research communities (e.g. dealing with human-centric dimensions) so that the empirical software engineering community does not only preach to the choir?
\item How can and should future software engineering researchers be educated?
\end{compactenum}

Despite all well-meant attempts so far to implement inter- and trans-disciplinary training and evaluation schemes, these are still largely oriented towards specific disciplinary standards~\cite{Brown_2015}. Given the high diversity of disciplinary professional backgrounds in empirical software engineering, it is often difficult if not impossible to find a common ground on the principles and terms established so far. At the same time, the young age of our field also creates a unique chance to enhance the heterogenous approaches in empirical software engineering we have in place ranging from in-depth case studies to large scale statistical approaches in the various software application domains.

\section{Conclusion}

Software engineering research is more and more confronted with challenging questions that increase the need for interdisciplinary research. This especially holds when considering the human-centric facets in our field. Although we can already observe an increase in the call for more interdisciplinary research and studies investigating interdisciplinary topics, \emph{interdisciplinary research} comes with additional challenges barely discussed from a scientific (epistemological) point of view.

In this position statement, we have reflected upon the evolution and epistemological setting of empirical software engineering in the attempt to position itself as a scientific practice. Based on the lessons from the history of science, being full of controversial and in parts competing movements, and reflecting social mechanisms of in parts competing research communities, we have:
\begin{compactenum}
\item Elaborated a pragmatic view on empirical research for software engineering that recognises not a gold standard for scientific practice, but that reflects its cyclic nature in an attempt to undertaking research relying on various forms of methods, data sources, and theories from various disciplines.
\item Motivated a pragmatic path towards a balanced combination of methods and domain expertise via flexible, symmetrical, interdisciplinary research.
\item Adopted five rules of thumb from other, more experienced fields of deep interdisciplinary collaborations and pointed to new challenges.
\end{compactenum}

We argued, in particular, for the need for \emph{symmetrical coalitions} between empirical software engineering researchers and researchers from other evidence-based disciplines to shift from treating empirical software engineering as one singular developing discipline moving towards a paradigmatic stage of normal science, but as an \emph{interdiscipline}. In such an interdiscipline, social, cultural, and human-centric issues shape the symmetrical configurations of questions, research methods, and teams, which pose, however, new challenges on research practice and higher education. While we cannot propose quick solutions to these challenges, we postulate the need for a richer discussion of these in our research community and beyond. We know that such collaborations and approaches take time, effort (and funding) and that every attempt to open up spaces for symmetrical collaborations also results in trade-offs in terms of disciplinary standards. But we believe that we will otherwise not be able to cope with today's complex social, cultural and human-centric questions in software engineering in the long run. Our hope is that with this manuscript, we therefore not only oppose the stubborn voices who are still generally critical towards empiricism, but that we add to the voices of those defending the values of empirical research in software engineering in general and interdisciplinary research in particular. 

\section*{Acknowledgements}
We would like to thank Magne J{\o}rgensen for stimulating and fruitful discussions on this important topic and Daniel Graziotin as well as the anonymous reviewers of the Journal of Systems and Software for their constructive and profound feedback on earlier versions of this manuscript.

\bibliographystyle{model1-num-names}
\bibliography{Literature}

\section*{Authors}
\textbf{Daniel M\'{e}ndez Fern\'{a}ndez} is a senior lecturer in software and systems engineering at the Technical University of Munich, Germany, and director of the junior research groups at the Centre Digitisation.Bavaria, an interdisciplinary research incubator for topics related to the software-driven digital transformation. His research is on evidence-based, human-centred software engineering with a particular focus on interdisciplinary, qualitative research in Requirements Engineering and quality management. He has regularly published in various software engineering venues and has occupied several key positions in venues of the empirical software engineering community. He is further a member of the ACM, the IEEE Computer Society, and the German association of university professors and lecturers, and he serves as the University representative to ISERN, the International Empirical Software Engineering Research Network.
\newline

\textbf{Jan-Hendrik Passoth} is head of the Digital Media Lab at the Munich Center for Technology in Society at the Technical University of Munich, Germany. As a trained sociologist, his research in science and technology studies (STS) focuses on the social and political role of data and algorithms and developing strategies and tactics for a critical and engaged design of digital technologies in collaboration with software developers and practitioners. Projects like "Media Centers of the Future" with the Bayerische Rundfunk or the EU H2020 Project IMPROVE in the field of industrial automation combine qualitative methods (ethnography, focus groups, etc.) with inventive approaches and digital methods. He is a member of the European Association for the Studies of Science and Technology (EASST) and the Society for Social Studies of Science (4S).

\end{document}